\documentstyle[proceedings,numreferences]{crckapbk}
\begin{opening}
\title{GRAVITATIONAL WAVES FROM COMPACT BODIES}
\author{KIP S. THORNE}
\institute{California Institute of Technology \\
Pasadena, California USA}
\end{opening}
\begin{document}

\def\agt{
\mathrel{\raise.3ex\hbox{$>$}\mkern-14mu\lower0.6ex\hbox{$\sim$}}
}
\def\alt{
\mathrel{\raise.3ex\hbox{$<$}\mkern-14mu\lower0.6ex\hbox{$\sim$}}
}

\bibliographystyle{unsrt}    

\section{Introduction}\label{introduction}

\null\footnote{This paper will
be published in {\it Proceedings of IAU Symposium 165, Compact Stars in
Binaries,} edited by J. van Paradijs, E. van den Heuvel, and E. Kuulkers
(Kluwer Academic Publishers).}
According to general relativity theory, compact concentrations of energy
(e.g., neutron stars and black holes) should warp spacetime strongly, and
whenever such an energy concentration changes shape,
it should create a dynamically changing
spacetime warpage that propagates out
through the Universe at the speed of light.
This propagating warpage is called {\it gravitational
radiation}---a name that arises from general relativity's description of
gravity as a consequence of spacetime warpage.

There are a number of efforts, worldwide, to detect gravitational
radiation.  These efforts are driven in part by the desire to
``see gravitational waves in the flesh,'' but more importantly by
the goal of using the
waves as a probe of the Universe and of the nature of gravity.  They
should be a powerful probe, since they carry very detailed
information about gravity and their sources.

In this lecture I shall describe the prospects to study gravitational
waves from astronomical systems that exist in our Universe today.
Such systems are expected to radiate in the ``high-frequency''
($1$---$10^4$Hz) and ``low-frequency'' ($10^{-4}$---$1$Hz)
gravitational-wave bands.  By contrast, gravitational waves from the
early universe (which I shall not discuss here) should populate a far
wider band of frequencies, $\sim 10^{-18}$Hz---$10^{+8}$Hz.

The high-frequency band is the domain of Earth-based gravitational-wave
detectors: resonant-mass antennas, and most especially laser interferometers.
A number of interesting compact sources fall in this
band:  the stellar collapse to a neutron star or black
hole in our Galaxy and distant galaxies, which triggers supernovae; the
rotation and vibration of neutron
stars in our Galaxy; and the coalescence of neutron-star and
stellar-mass black-hole binaries ($M \alt 1000 M_\odot$)
in distant galaxies.

The low-frequency band is the domain of detectors
flown in space (in Earth orbit or in interplanetary orbit). The most
important of these are the Doppler tracking of spacecraft via
microwave signals sent
from Earth to the spacecraft and there transponded back to Earth (a
technique that NASA has pursued since the early 1970's), and optical
tracking of spacecraft by each other (laser interferometry in space, a
technique now under development for possible flight in $\sim 2014$).
The low-frequency band should be populated by waves from short-period
binary stars in our own Galaxy (main-sequence binaries, cataclysmic
variables, white-dwarf binaries, neutron-star binaries, ...); by waves
from white dwarfs, neutron stars, and small black holes spiraling into
massive black holes ($M \sim 3\times 10^5$ to $3\times 10^7 M_{\odot}$)
in distant
galaxies; and by waves from the inspiral and coalescence of massive
black-hole binaries ($M\sim100$ to $10^8 M_\odot$) in distant galaxies.

\begin{figure}
\vskip 9.5pc
\special{hscale=66 vscale=66 hoffset=-2 voffset=-13
psfile=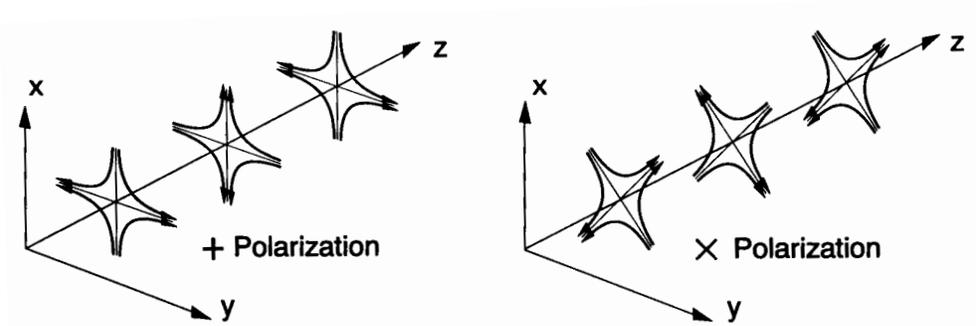}
\caption{The lines of force associated with the two polarizations of
a gravitational wave.  (From Ref. \protect\cite{ligoscience}.)
}
\label{fig:forcelines}
\end{figure}

The body of this lecture consists of four principal sections.
Section 2 describes a network of high-frequency, ground-based
laser interferometers that is currently under construction, and then
Sec.\ 3 describes the
high-frequency sources the network will search for, and the information that
we hope to glean from
their waves.  Section 4 describes a system of low-frequency,
space-based interferometers planned for launch in $\sim 2014$, and
Sec.\ 5 describes the low-frequency sources they will search for
and the information carried by their waves.

\section{Ground-Based Laser Interferometers}\label{gbint}

\subsection{Wave Polarizations, Waveforms, and How an Interferometer
Works}

According to general relativity theory, a gravitational wave has two
linear polarizations,
conventionally called $+$ (plus) and $\times$ (cross).  Associated with
each polarization there is a gravitational-wave field, $h_+$ or $h_\times$,
which oscillates in time and propagates with the speed of light.  Each
wave field produces tidal forces (gravitational
stretching and squeezing forces) on
any object or detector through which it passes.  If the object is small
compared to the waves' wavelength (as is the case for ground-based
interferometers and resonant mass antennas), then relative to the
object's center, the forces have the quadrupolar patterns shown in Figure
\ref{fig:forcelines}.
The names ``plus'' and ``cross'' are derived from the orientations of the
axes that characterize the force patterns \cite{300yrs}.

\begin{figure}
\center
\vskip 11.7pc
\special{hscale=60 vscale=60 hoffset=27 voffset=-5
psfile=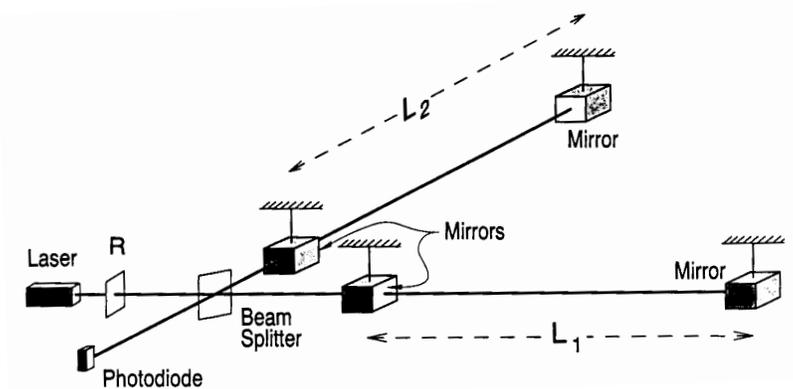}
\caption{Schematic diagram of a laser interferometer
gravitational wave detector.  (From Ref.\ \protect\cite{ligoscience}.)
}
\label{fig:interferometer}
\end{figure}

A laser interferometer gravitational wave detector (``interferometer'' for
short) consists of four masses that hang from vibration-isolated
supports as shown in Figure \ref{fig:interferometer}, and the indicated
optical system for
monitoring the separations between
the masses \cite{300yrs,ligoscience}. Two masses are near
each other, at the corner of an ``L'',
and one mass is at the end of each of the L's long arms.  The arm
lengths are nearly equal, $L_1 \simeq L_2 = L$.  When a gravitational
wave, with frequencies high compared to the masses' $\sim 1$ Hz pendulum
frequency, passes through the detector, it pushes the masses back and
forth relative to each other as though they were free from their
suspension wires, thereby changing the arm-length difference,
$\Delta L \equiv L_1-L_2$.  That change is
monitored by laser interferometry in such a way that the output of the
photodiode (the interferometer's output) is directly proportional to
$\Delta L(t)$.

If the gravitational waves are coming from overhead or
underfoot
and the axes of the $+$ polarization coincide with the arms'
directions, then it is the waves' $+$ polarization that drives the masses, and
$\Delta L(t) / L = h_+(t)$.  More generally,
the interferometer's output is a linear combination of the two
wave fields:
\begin{equation}
{\Delta L(t)\over L} = F_+h_+(t) + F_\times h_\times (t) \equiv h(t)\;.
\label{dll}
\end{equation}
The coefficients $F_+$ and $F_\times$ are of order unity and depend in a
quadrupolar manner on
the direction to the source and the orientation of the detector \cite{300yrs}.
The combination $h(t)$ of the two waveforms is called the
{\it gravitational-wave strain} that acts on the detector; and the time
evolutions of $h(t)$, $h_+(t)$, and $h_\times(t)$ are sometimes called
{\it waveforms}.

\subsection{Wave Strengths and Interferometer Arm Lengths}

The strengths of the waves from a gravitational-wave source can be
estimated using the
``Newtonian/quadrupole'' approximation to the Einstein field equations.
This approximation says that $h\simeq (G/c^4)\ddot Q/r$, where
$\ddot Q$ is the second time derivative of the source's quadrupole
moment, $r$ is the distance of the source from Earth, and $G$ and $c$
are Newton's gravitation constant and the speed of light.
The strongest sources will be highly nonspherical and thus will
have $Q\simeq ML^2$, where $M$ is their mass and $L$ their size, and
correspondingly will have $\ddot Q \simeq 2Mv^2 \simeq 4 E_{\rm
kin}^{\rm ns}$, where $v$ is their internal velocity and
$E_{\rm kin}^{\rm ns}$ is the nonspherical part of their
internal kinetic energy.  This provides us with the estimate
\begin{equation}
h\sim {1\over c^2}{4G(E_{\rm kin}^{\rm ns}/c^2) \over r}\;;
\label{hom}
\end{equation}
i.e., $h$ is about 4 times the gravitational potential produced at Earth by
the mass-equivalent of the source's nonspherical, internal kinetic
energy---made dimensionless by dividing by $c^2$.  Thus, in order to
radiate strongly, the source must have a very large, nonspherical,
internal kinetic energy.

The best known way to achieve a huge internal kinetic energy is via
gravity; and by energy conservation (or the virial theorem), any
gravitationally-induced kinetic energy must be of order the source's
gravitational potential energy.  A huge potential energy, in turn,
requires that the source be very compact, not much larger than its own
gravitational radius.  Thus, the strongest gravity-wave sources must be
highly compact, dynamical concentrations of large amounts of mass (e.g.,
colliding and coalescing black holes and neutron stars).

Such sources cannot remain highly dynamical for long; their motions will
be stopped by energy loss to gravitational waves and/or the formation of
an all-encompassing black hole.  Thus, the strongest sources should be
transient.  Moreover, they should be very rare --- so rare that to see a
reasonable event rate will require reaching out through a substantial
fraction of the Universe.  Thus, just as the strongest radio waves
arriving at Earth
tend to be extragalactic, so also the strongest gravitational waves are
likely to be extragalactic.

For highly compact, dynamical objects that radiate in the high-frequency band,
e.g.\ colliding and coalescing neutron stars and stellar-mass black
holes, the internal, nonspherical kinetic energy
$E_{\rm kin}^{\rm ns} /c^2$ is of order the mass of
the Sun; and, correspondingly, Eq.\ (\ref{hom}) gives $h\sim 10^{-22}$
for such sources at the Hubble
distance,
$h\sim 10^{-21}$ at 200 Mpc (a best-guess distance for several
neutron-star coalescences per year; Sec.\ \ref{coalescencerates}),
and $h\sim 10^{-20}$ at the
Virgo cluster (15 Mpc). These numbers set the
scale of sensitivities that ground-based interferometers seek to achieve:
$h \alt 10^{-21}$ to $10^{-22}$.

When one examines the technology of laser interferometry, one sees good
prospects to achieve measurement accuracies $\Delta L \sim 10^{-16}$ cm
(1/1000 the diameter of the nucleus of an atom).  With such an accuracy,
an interferometer must have an arm length $L = \Delta L/h \sim 1$ to 10
km, in order to achieve the desired wave sensitivities, $10^{-21}$ to
$10^{-22}$.  This sets the scale of the interferometers that are now
under construction.

\subsection{LIGO, VIRGO, and the International Interferometric Network}
\label{network}

Interferometers are plagued by non-Gaussian noise, e.g.\ due
to sudden strain releases in the wires that suspend the masses.  This
noise prevents a single interferometer, by itself, from detecting with
confidence short-duration gravitational-wave bursts (though it might
be possible for a single interferometer to search for the periodic
waves from known pulsars).   The non-Gaussian noise can be removed
by cross correlating two, or preferably three or more, interferometers
that are networked together at widely separated sites.

The technology and techniques for such interferometers have been under
development for nearly 25 years, and plans for km-scale
interferometers have been developed over the past 13 years.  An
international network consisting of three km-scale interferometers, at
three widely separated sites, is now in the early stages of
construction. It includes two sites of
the American LIGO Project (``Laser Interferometer Gravitational Wave
Observatory'') \cite{ligoscience}, and one site of the French/Italian VIRGO
Project (named after the Virgo cluster of galaxies) \cite{virgo}.

LIGO will consist of two
vacuum facilities with 4-kilometer-long arms, one in Hanford, Washington
(in the northwestern United States)
and the other in Livingston, Louisiana (in the southeastern United States).
These facilities are designed to house
many successive generations of interferometers without the necessity of
any major facilities upgrade; and after a planned
future expansion,
they will be able to house several interferometers at once, each
with a different optical configuration optimized for a different type of
wave (e.g., broad-band burst, or narrow-band periodic wave, or
stochastic wave).  The LIGO facilities and their first interferometers
are being constructed by a team of about 80 physicists and engineers at
Caltech and MIT, led by Barry Barish (the PI) and Gary Sanders (the
Project Manager).  Substantial contributions are also being made by
scientists at other institutions.

The VIRGO Project is building one vacuum facility in Pisa, Italy, with
3-kilometer-long arms.  This facility and its first interferometers are
a collaboration of more than a hundred physicists and engineers
at the INFN (Frascati, Napoli, Perugia, Pisa), LAL (Orsay), LAPP
(Annecy), LOA (Palaiseau), IPN (Lyon), ESPCI (Paris), and the University
of Illinois (Urbana), under the leadership of Alain Brillet and
Adalberto Giazotto.

Both LIGO and VIRGO are scheduled for completion
in the late 1990s, and their first gravitational-wave searches are
likely to be performed in 2000 or 2001.

LIGO alone, with its two sites which have parallel arms, will be able to
detect an incoming gravitational wave, measure one of its two waveforms,
and (from the time delay between the two sites) locate its source to within a
$\sim 1^{\rm o}$ wide annulus on the sky.
LIGO and VIRGO together, operating as a {\sl coordinated international
network}, will be able to locate the source
(via time delays plus the interferometers' beam patterns) to within a
2-dimensional error box with size
between several tens of arcminutes and several degrees, depending on
the source direction and on
the amount of high-frequency structure in the waveforms.
They will also be able to monitor both waveforms $h_+(t)$ and
$h_\times(t)$ (except for frequency components above about 1kHz
and below about 10 Hz, where the interferometers' noise becomes severe).

The accuracies of the direction measurements and the ability to
monitor more than one
waveform will be severely compromised when the source lies anywhere near
the plane formed by the three LIGO/VIRGO interferometer locations.  To
get good all-sky coverage will require a fourth interferometer at a site
far out of that plane; Japan and Australia would be excellent locations,
and research groups there are carrying out research and development on
interferometric detectors, aimed at such a possibility.  A 300-meter
prototype interferometer called TAMA is under construction in Tokyo,
and a 400-meter prototype called AIGO400 has been proposed for
construction north of Perth.

Two other groups are major players in this field, one in Britain led
by James Hough, the other in Germany, led by Karsten Danzmann. These
groups each
have two decades of experience with
prototype interferometers (comparable experience to the LIGO team
and far more than anyone else) and great expertise.  Frustrated by
inadequate financing for
a kilometer-scale interferometer, they
are constructing, instead, a 600 meter
system called GEO600 near Hannover, Germany.  Their goal is to develop,
from the outset, an interferometer with the sort of advanced design that
LIGO and VIRGO will attempt only as a ``second-generation''
instrument, and thereby achieve sufficient sensitivity
to be full partners in the international
network's first gravitational-wave searches; they then would offer a
variant of their
interferometer as a candidate for second-generation operation in the
much longer arms of LIGO
and/or VIRGO.  It is a seemingly audacious plan, but with their
extensive experience and expertise, the British/German collaboration
might pull it off successfully.

\section{High-Frequency Gravitational-Wave Sources}

\subsection{Coalescing Compact Binaries}
\label{coalescing_binaries}

The best understood of all gravitational-wave sources are coalescing,
compact binaries composed of neutron stars (NS) and black holes (BH).
These NS/NS, NS/BH, and BH/BH binaries may well become the ``bread and
butter'' of the LIGO/VIRGO diet.

The Hulse-Taylor \cite{hulse_taylor,taylor} binary pulsar,
PSR 1913+16, is an example of a NS/NS binary
whose waves could be measured by LIGO/VIRGO, if we were to wait long
enough.  At present PSR1913+16 has an orbital frequency of about 1/(8 hours)
and emits its waves predominantly at twice this frequency, roughly
$10^{-4}$ Hz, which is in the low-frequency band---far too low to be
detected by LIGO/VIRGO.  However,
as a result of their loss of orbital energy to gravitational waves,
the PSR1913+16 NS's are gradually spiraling inward.  If we wait roughly
$10^8$ years, this inspiral will bring the waves into the LIGO/VIRGO
high-frequency band. As the NS's continue their inspiral,
the waves will then sweep upward in frequency, over a time of about 15
minutes, from 10 Hz to $\sim 10^3$ Hz, at which point the NS's will
collide and coalesce.  It is this last 15 minutes
of inspiral, with $\sim 16,000$
cycles of waveform oscillation, and the final coalescence,
that LIGO/VIRGO seeks to monitor.

\begin{figure}
\center
\vskip23.2pc
\special{hscale=60 vscale=60 hoffset=8 voffset=-5
psfile=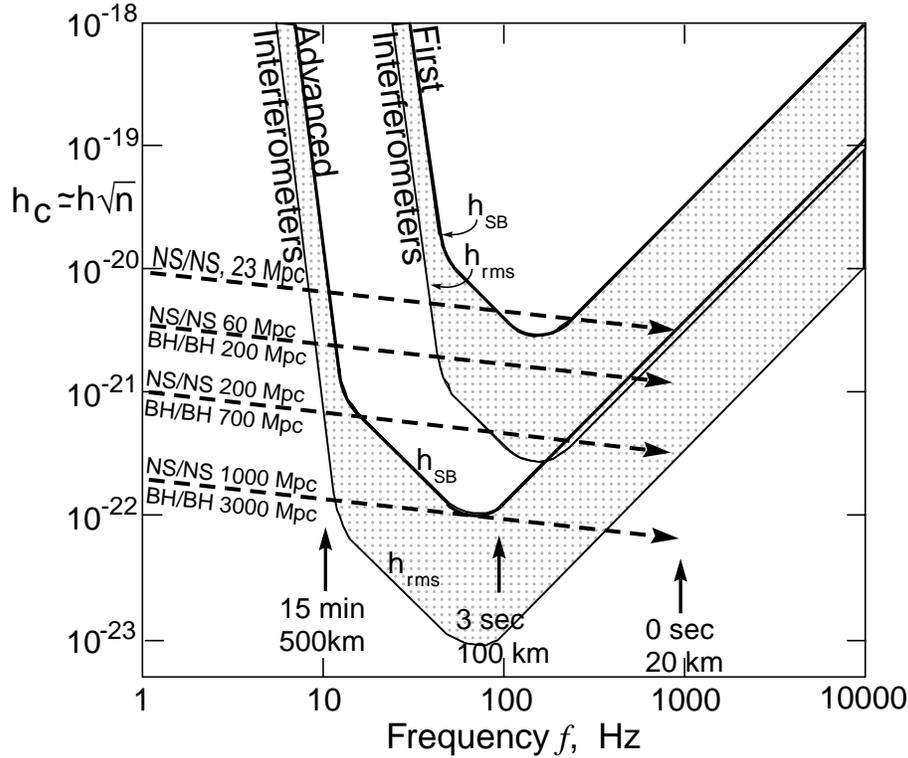}
\caption{LIGO's projected broad-band noise $h_{\rm rms}$ and sensitivity
to bursts $h_{\rm SB}$
\protect\cite{ligoscience} compared with the
strengths of the
waves from the last few minutes of inspiral of compact binaries.  The
signal to noise ratios are
$\protect\sqrt 2$ higher than in Ref.\ \protect\cite{ligoscience}
because of a factor 2 error in Eq.~(29) of Ref.\ \protect\cite{300yrs}.
}
\label{fig:CBStrengthSensitivity}
\end{figure}

\subsubsection{Wave Strengths Compared to LIGO Sensitivities}

Figure \ref{fig:CBStrengthSensitivity}
compares the projected sensitivities of interferometers in
LIGO \cite{ligoscience}
with the wave strengths from the last few minutes of inspiral
of BH/BH, NS/BH, and NS/NS binaries at various distances from Earth.  The
two solid
curves at the bottoms of the stippled regions (labeled $h_{\rm rms}$)
are the rms noise levels
for broad-band waves that have optimal direction and polarization.
The tops of the stippled regions (labeled $h_{\rm SB}$ for ``sensitivity to
bursts'') are the sensitivities for highly confident
detection of randomly polarized, broad-band waves from random directions
(i.e., the sensitivities for high confidence that any such observed
signal is not a false alarm due to Gaussian noise).  The upper stippled
region and its bounding curves
are the expected performances of the first interferometers in LIGO; the
lower stippled region and curves are performances of more advanced
LIGO interferometers.

As the NS's and/or BH's spiral inward, their waves
sweep upward in frequency (left to right in the diagram).  The dashed
lines show their ``characteristic'' signal strength $h_c$
(approximately the amplitude $h$ of the waves' oscillations
multiplied by the square root of the number of cycles spent near a given
frequency, $\sqrt n$); the signal-to-noise ratio is this $h_c$
divided by the detector's $\sqrt5 h_{\rm rms}$, $S/N = h_c/(\sqrt5
h_{\rm rms})$, where the $\sqrt5$
converts $h_{\rm rms}$ from ``optimal direction and polarization'' to
``random direction and polarization'') \cite{ligoscience,300yrs}.  The arrows
along the NS/NS inspiral track indicate the time until
final coalescence and the separation between the NS centers
of mass.  Each NS is assumed to have a mass of 1.4 suns and a radius
$\sim 10$ km; each BH, 10 suns and $\sim 20$ km.

Notice that the signal strengths in Fig.\ \ref{fig:CBStrengthSensitivity}
are in good accord with rough estimates based on Eq.\ (\ref{hom});
at the endpoint (right end) of each inspiral, the number of cycles $n$ spent
near that frequency is of order unity, so the quantity plotted, $h_c
\simeq h\sqrt
n$, is about equal to $h$---and at distance 200 Mpc is about $10^{-21}$,
as Eq.\ (\ref{hom}) predicts.

\subsubsection{Coalescence Rates}
\label{coalescencerates}

Such final coalescences are few and far between in our own galaxy:
about one every 100,000 years, according to 1991
estimates by Phinney \cite{phinney} and
by Narayan, Piran, and Shemi \cite{narayan},
based on the statistics of binary pulsar searches
in our galaxy which found three that will coalesce in less than
$10^{10}$ years.  Extrapolating out through the universe
on the basis of the density of
production of blue light (the color produced predominantly by
massive stars), Phinney \cite{phinney} and Narayan et.\ al.\ \cite{narayan}
infer that to see several
NS/NS coalescences per year, LIGO/VIRGO will have to look out to a
distance of about 200 Mpc (give or take a factor $\sim 2$); cf.\ the
``NS/NS inspiral, 200 Mpc'' line in Fig.\ \ref{fig:CBStrengthSensitivity}.
Since these estimates were made, the binary pulsar searches have been
extended through a significantly larger volume of the galaxy than
before, and no new ones with coalescence times $\alt 10^{10}$ years have
been found; as a result, the binary-pulsar-search-based
best estimate of the coalescence rate should be
revised downward \cite{bailes}, perhaps to as little as one every
million years in our
galaxy, corresponding to a distance $400$ Mpc for several per year
\cite{bailes}.

A rate of one every million years
in our galaxy is $\sim 1000$ times smaller than the birth rate of
the NS/NS binaries' progenitors:
massive, compact, main-sequence binaries \cite{phinney,narayan}.
Therefore, either 99.9 per cent of progenitors
fail to make it to the NS/NS state (e.g., because of binary disruption
during a supernova or forming T\.ZO's),
or else they do make it, but they wind up as a
class of NS/NS binaries that has not yet been discovered in any of the
pulsar searches.  Several experts on binary evolution have argued for
the latter \cite{tutukov_yungelson,yamaoka,vandenheuvel,lipunov}: most
NS/NS binaries, they suggest, may form with such short orbital periods
that their lifetimes to coalescence are significantly shorter than
normal pulsar lifetimes ($\sim 10^7$ years); and with such short
lifetimes, they have been missed in pulsar searches.  By modeling the
evolution of the galaxy's binary star population, the binary experts
arrive at
best estimates as high as $3\times 10^{-4}$ coalescences per year in our
galaxy, corresponding to several per year out to 60 Mpc distance
\cite{tutukov_yungelson}.  Phinney \cite{phinney} describes other
plausible populations of NS/NS binaries that could increase the event
rate, and he argues for ``ultraconservative'' lower and upper limits of
23 Mpc and 1000Mpc
for how far one must look to see several coalescence per year.

By comparing these rate estimates with the signal strengths in Fig.\
\ref{fig:CBStrengthSensitivity}, we see that (i) the first interferometers
in LIGO/VIRGO have a moderate but not high probability of seeing
NS/NS coalescences; (ii) advanced interferometers are almost certain of
seeing them (the requirement that
this be so was one factor that forced the LIGO/VIRGO arm lengths to be
so long, several kilometers); and (iii) they are most likely to be
discovered roughly half-way between the first and advanced
interferometers---which means by an improved variant of the first
interferometers several years after LIGO operations begin.

We have no good observational handle on the coalescence rate of NS/BH or
BH/BH binaries.  However, theory suggests that their progenitors might
not disrupt during the stellar collapses that produce the NS's and BH's,
so their coalescence rate could be about the same as the birth
rate for their progenitors: $\sim 1/100,000$ years in our galaxy.  This
suggests that within 200 Mpc distance there might be several NS/BH or
BH/BH coalescences per year.
\cite{phinney,narayan,tutukov_yungelson,lipunov}.
This estimate should be regarded as a
plausible upper limit on the event rate and lower limit on the distance to
look \cite{phinney,narayan}.

If this estimate is correct, then NS/BH and BH/BH binaries will be seen
before NS/NS, and might be seen by the first LIGO/VIRGO interferometers
or soon thereafter; cf.\ Fig.  \ref{fig:CBStrengthSensitivity}.
However, this estimate is far less certain than the
(rather uncertain) NS/NS estimates!

Once coalescence waves have been discovered, each further improvement of
sensitivity by a factor 2 will increase the event rate by $2^3 \simeq
10$.  Assuming a rate of several NS/NS per year at 200 Mpc, the advanced
interferometers of Fig.\ \ref{fig:CBStrengthSensitivity} should see
$\sim 100$ per year.

\subsubsection{Inspiral Waveforms and the Information They Can Bring}

Neutron stars and black holes have such intense self gravity that it is
exceedingly difficult to deform them.  Correspondingly, as they spiral
inward in a compact binary, they do not gravitationally deform each other
significantly until several orbits before their final
coalescence \cite {kochanek,bildsten_cutler}.  This means
that the inspiral waveforms are determined to high accuracy by
only a few, clean parameters:
the masses and spin angular momenta of the bodies, and the initial
orbital elements (i.e.\ the elements when the waves enter the LIGO/VIRGO band).

\begin{figure}
\vskip 14.0pc
\special{hscale=62 vscale=62 hoffset=-2 voffset=-13
psfile=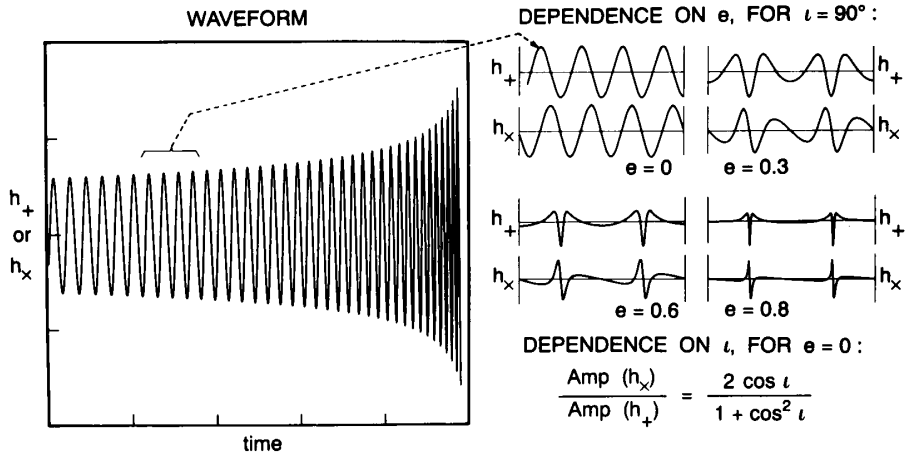}
\caption{Waveforms from the inspiral of a compact binary, computed using
Newtonian gravity for the orbital evolution and the quadrupole moment
approximation for the wave generation.  (From Ref.\
\protect\cite{ligoscience}.)}
\label{fig:NewtonInspiral}
\end{figure}

Though tidal deformations are negligible during inspiral, relativistic
effects can be very important.
If, for the moment, we ignore the relativistic effects---i.e., if we
approximate gravity as Newtonian and the wave generation as due to the
binary's oscillating quadrupole moment \cite{300yrs},
then the shapes of the inspiral
waveforms $h_+(t)$ and $h_\times(t)$
are as shown in Fig.\ \ref{fig:NewtonInspiral}.

The left-hand graph in Fig.\ \ref{fig:NewtonInspiral} shows the waveform
increasing in
amplitude and sweeping upward in frequency
(i.e., undergoing a ``chirp'')
as the binary's bodies spiral closer and closer together.  The ratio of
the amplitudes
of the two polarizations is determined by the inclination $\iota$ of the
orbit to our line of sight (lower right in Fig.\ \ref{fig:NewtonInspiral}).
The
shapes of the individual waves, i.e.\ the waves' harmonic content, are
determined by the orbital eccentricity (upper right).  (Binaries
produced by normal stellar evolution should be highly circular due to
past radiation reaction forces, but compact
binaries that form by capture events, in dense star clusters that might
reside in galactic nuclei \cite{quinlan_shapiro}, could be quite
eccentric.)  If, for simplicity, the
orbit is circular, then the rate at which
the frequency sweeps or ``chirps'', $df/dt$
[or equivalently the number of cycles
spent near a given frequency, $n=f^2(df/dt)^{-1}$] is determined solely, in the
Newtonian/quadrupole approximation, by the binary's so-called {\it
chirp mass}, $M_c \equiv (M_1M_2)^{3/5}/(M_1+M_2)^{1/5}$ (where $M_1$
and $M_2$ are the two bodies' masses).
The amplitudes of the two waveforms are determined by the chirp mass,
the distance to the source, and the orbital inclination.  Thus
(in the Newtonian/quadrupole
approximation), by measuring the two amplitudes, the frequency sweep, and
the harmonic content of the inspiral waves, one can determine as direct,
resulting observables, the source's distance, chirp mass, inclination,
and eccentricity \cite{schutz_nature86,schutz_grg89}.

\begin{figure}
\vskip18pc
\special{hscale=140 vscale=140 hoffset=-12 voffset=-5
psfile=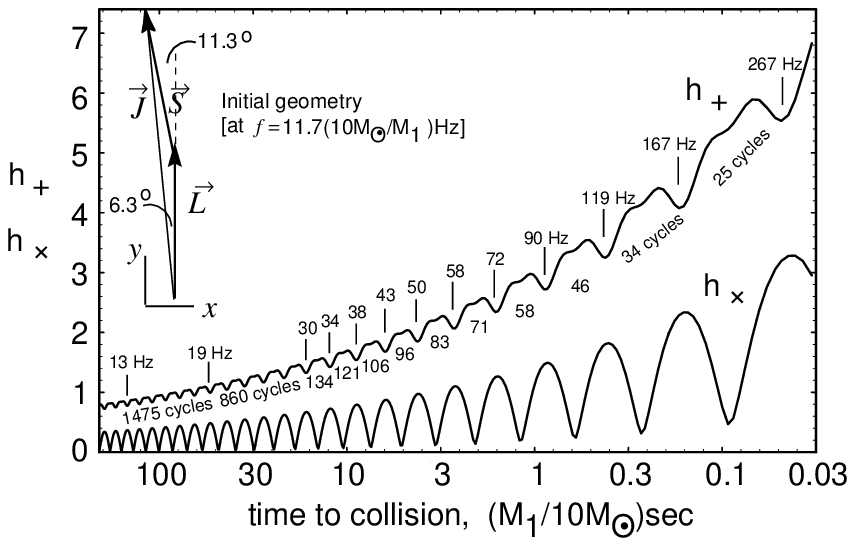}
\caption{Modulational envelope for the waveform from
a $1 M_\odot$ nonspinning NS spiraling into a $10 M_\odot$, rapidly
spinning Kerr black hole (spin parameter $a = 1$).  The orbital angular
momentum $\bf L$ is inclined by $\alpha = 11.3$ degrees to the hole's
spin angular momentum $\bf S$, and the two precess around
${\bf J} ={\bf L} + {\bf S}$, whose direction remains
fixed in space as $L = |{\bf L}|$ shrinks and $S = |{\bf S}| = M_{\rm
BH}a$ remains constant.
The precession
modulates the waves by an amount that depends on
(i) the direction to Earth (here along the
initial $\bf L \times \bf S$, i.e. out of the paper)
and (ii) the
orientation of the detector's arms (here parallel to the figure's
initial $\bf L$ and to ${\bf L}\times$(direction to Earth)
for $h_+$, and rotated 45 degrees for $h_\times$).
The figure shows the waveforms' modulational envelopes (in
arbitrary units, the same for $h_+$ and $h_\times$),
parametrized by the wave frequency $f$ and the number of cycles
of {\it oscillation} between the indicated $f$'s.  The total
number of {\it precessions} from
$f$ to coalescence is $N_{\rm prec} \simeq (5/64\pi)(Ma/\mu)(\pi
Mf)^{-2/3} \simeq 20(f/10\hbox{Hz})^{-2/3}$.
(From \protect\cite{last3minutes,precess}.)
}
\label{fig:precess}
\end{figure}

As in binary pulsar observations \cite{taylor},
so also here, relativistic effects add further information:
they influence the rate of frequency sweep and produce waveform
modulations in ways that
depend on the binary's dimensionless ratio $\eta = \mu/M$ of reduced mass
$\mu = M_1 M_2/(M_1 + M_2)$ to total mass $M = M_1 + M_2$ \cite{lincoln_will}
and on the spins of the binary's two bodies \cite{kidder_will_wiseman};
These relativistic effects are reviewed and discussed at length in
Refs.\ \cite{last3minutes,will_nishinomiya}.  Two deserve special
mention: (i) As the waves emerge from the binary, some of them get
backscattered one or more times off the binary's spacetime curvature,
producing wave {\it tails}.  These tails act back on the binary,
modifying its inspiral rate in a measurable way.  (ii)  If the orbital
plane is inclined to one or both of the binary's spins, then
the spins drag inertial frames in the binary's
vicinity (the ``Lense-Thirring effect''), this frame dragging causes
the orbit to precess, and the precession modulates the
waveforms \cite{last3minutes,precess,kidder}.  Figure
\ref{fig:precess} shows the resulting modulation for a $1 M_\odot$
NS spiraling into a rapidly spinning, $10 M_\odot$ BH.

Remarkably, the relativistic corrections to the
frequency sweep will be measurable with very high
accuracy, even though they are typically $\alt 10$ per cent of the
Newtonian contribution, and even though the
typical signal to noise ratio will be only $\sim 9$ even after optimal
signal processing.  The reason is as
follows \cite{cutler_flanagan,finn_chernoff,last3minutes}:

The frequency sweep will be monitored by the method of ``matched
filters''; in other words, the incoming, noisy signal will be cross
correlated with theoretical templates.  If the signal and the templates
gradually
get out of phase with each other by more than $\sim 1/10$ cycle as the
waves sweep
through the LIGO/VIRGO band, their cross correlation will be significantly
reduced.
Since the total number of cycles spent in the LIGO/VIRGO band will
be $\sim 16,000$ for a NS/NS binary, $\sim 3500$ for NS/BH, and $\sim
600$ for BH/BH, this means that LIGO/VIRGO should be able to measure the
frequency sweep to a fractional precision $\alt 10^{-4}$,
compared to which the relativistic effects
are very large.  (This is essentially the same method as
Joseph Taylor and colleagues use for high-accuracy radio-wave measurements of
relativistic effects in binary pulsars \cite{taylor}.)

Preliminary analyses, using the theory of optimal signal processing,
predict the following typical accuracies for LIGO/VIRGO measurements
based solely on the frequency sweep (i.e., ignoring modulational
information)
\cite{poisson_will,cutler_flanagan,finn_chernoff,jaronowski_krolak,last3minutes}: (i) The
chirp mass $M_c$
will typically be measured, from the Newtonian part of the frequency
sweep, to $\sim 0.04\%$ for a NS/NS binary and
$\sim 0.3\%$ for a system containing at least one BH.
(ii) {\it If} we
are confident (e.g., on a statistical basis from measurements of many
previous binaries) that the spins are a few percent or less
of the maximum physically allowed, then the reduced mass $\mu$
will be measured to
$\sim 1\%$ for NS/NS and NS/BH binaries, and
$\sim 3\%$ for BH/BH binaries.  (Here and below NS means a
$\sim 1.4 M_\odot$
neutron star and BH means a $\sim 10 M_\odot$
black hole.) (iii) Because the
frequency dependences
of the (relativistic) $\mu$ effects
and spin effects are not
sufficiently different
to give a clean separation between $\mu$ and the spins,
if we have no prior knowledge of the spins, then
the spin$/\mu$ correlation will
worsen the typical accuracy of $\mu$ by a large factor,
to $\sim 30\%$ for NS/NS, $\sim 50\%$ for NS/BH, and
a factor $\sim 2$ for BH/BH \cite{poisson_will,cutler_flanagan}.
These worsened accuracies might be improved somewhat
by waveform modulations caused by the
spin-induced precession of the orbit \cite{precess,kidder},
and even without modulational information, a certain
combination of $\mu$ and the spins
will be determined to a few per cent.  Much
additional theoretical work is needed
to firm up the measurement accuracies.

To take full advantage of all the information in the inspiral waveforms
will require theoretical templates that are accurate, for given masses
and spins, to a fraction of a cycle during the entire sweep through the
LIGO/VIRGO band.  Such templates are being computed by an international
consortium of relativity theorists (Blanchet and Damour in France, Iyer
in India, Will and Wiseman in the U.S., and others)
\cite{will_nishinomiya,2pnresults}, using post-Newtonian expansions of
the Einstein field equations.  This enterprise is rather like computing
the Lamb shift to high order in powers of the fine structure
constant, for comparison with experiment.  The terms of leading order in
the mass ratio $\eta = \mu / M$ are being checked by a
Japanese-American consortium
(Poisson, Nakamura, Sasaki, Tagoshi, Tanaka) using the Teukolsky
formalism for weak perturbations of black holes
\cite{poisson_check,shibataetal}.  These small-$\eta$ calculations have
been carried to very high post-Newtonian order for circular orbits and no spins
\cite{nakamura_tagoshi,sasaki_tagoshi}, and from those results Cutler
and Flanagan \cite{cutler_flanagan1} have estimated the order to which the
full, finite-$\eta$ computations must be carried in order that
systematic errors in the theoretical templates will not significantly
impact the information extracted from the LIGO/VIRGO observational data.
The answer appears daunting: radiation-reaction effects must be computed
to three full post-Newtonian orders [six orders in $v/c =$(orbital
velocity)/(speed of light)] beyond the leading-order radiation reaction,
which itself is 5 orders in $v/c$ beyond the Newtonian theory of
gravity.

It is only about ten years since controversies over the leading-order
radiation reaction \cite{quadrupole_controversy} were resolved by a
combination of theoretical
techniques and binary pulsar observations.  Nobody dreamed then that
LIGO/VIRGO observations will require pushing post-Newtonian computations
onward from $O[(v/c)^5]$ to $O[(v/c)^{11}]$.  This requirement epitomizes
a major change in the field of relativity research: At last, 80 years
after Einstein formulated general relativity, experiment has become a
major driver for theoretical analyses.

Remarkably, the goal of $O[(v/c)^{11}]$ is achievable.  The most difficult
part of the computation, the radiation reaction, has been evaluated to
$O[(v/c)^9]$ beyond Newton by the French/Indian/American consortium
\cite{2pnresults}
and as of this writing, rumors have it that $O[(v/c)^{10}]$ is coming
under control.

These high-accuracy waveforms are needed only for extracting information
from the inspiral waves, after the waves have been discovered; they are
not needed for the discovery itself.  The discovery is best achieved
using a different family of theoretical waveform templates, one that
covers the space of potential waveforms
in a manner that minimizes computation time instead
of a manner that ties quantitatively into general relativity
theory \cite{last3minutes}.  Such templates are in the early stage of
development \cite{krolak_kokkotas_schafer,apostolatos,sathyaprakash}.

LIGO/VIRGO observations of compact binary inspiral have the potential to
bring us far more information than just binary masses and spins:
\begin{itemize}
\item
They can be used for high-precision tests of general relativity.  In
scalar-tensor theories (some of which are highly attractive alternatives
to general relativity \cite{damour_nordvedt}), radiation reaction
due to emission
of scalar waves places a unique signature on those waves that LIGO/VIRGO
would detect---a signature that can be searched for with high precision
\cite{will_scalartensor}.
\item
They can be used to measure the Hubble constant, deceleration
parameter, and cosmological constant
\cite{schutz_nature86,schutz_grg89,markovic,chernoff_finn}.  The keys to
such measurements are that (i) advanced interferometers in
LIGO/VIRGO will be able to see NS/NS
out to cosmological redshifts $z \sim 0.3$, and NS/BH out to $z
\sim 2$.  (ii) The direct observables that can be extracted
from the
observed waves include the source's luminosity distance $r_{\rm L}$ (measured
to accuracy $\sim 10$ per cent in a large fraction of cases), and its
direction on the sky (to accuracy $\sim 1$ square degree)---accuracies
good enough that only one or a few electromagnetically-observed
clusters of galaxies should fall within the 3-dimensional
gravitational error boxes, thereby giving promise to joint
gravitational/electromagnetic statistical studies.  (iii) Another direct
gravitational observable is $(1+z)M$
where $z$ is redshift and $M$ is any mass in the system (measured to the
accuracies quoted above). Since the masses of NS's in binaries seem to
cluster around $1.4 M_\odot$, measurements of $(1+z)M$ can provide a
handle on the redshift, even in the absence of electromagnetic aid.
\item
For a NS or small BH spiraling into a massive $\sim 50$ to $500 M_\odot$
BH, the inspiral waves will carry a ``map'' of the spacetime geometry
around the big hole---a map that can be used, e.g., to test the theorem
that ``a black hole has no hair'' \cite{ryan_finn_thorne};
cf.\ Sec.\ \ref{inspiral_waves} below.
\end{itemize}

\subsubsection{Coalescence Waveforms and their Information}
\label{coalescence_waves}

The waves from the binary's final coalescence can bring us new
types of information.

{\bf BH/BH Coalescence}: In the case of a BH/BH binary, the coalescence
will excite large-amplitude, highly nonlinear vibrations of spacetime
curvature near the coalescing black-hole horizons---a phenomenon of
which we have very little theoretical understanding today.  Especially
fascinating will be the case of two spinning black holes whose spins are
not aligned with each other or with the orbital angular momentum.  Each
of the three angular momentum vectors (two spins, one orbital) will drag
space in its vicinity into a tornado-like swirling motion---the general
relativistic ``dragging of inertial frames,'' so the binary is rather
like two tornados with orientations skewed to each other, embedded inside a
third, larger tornado with a third orientation.  The dynamical evolution of
such a complex configuration of spacetime warpage (as revealed by its
emitted waves) may well bring us
surprising new insights into relativistic gravity.  Moreover, if the sum
of the BH masses is fairly large, $\sim 40$ to $200 M_\odot$, then the waves
should come off in a frequency range $f\sim 40$ to $200$ Hz where the
LIGO/VIRGO broad-band interferometers
have their best sensitivity and can best
extract
the information the waves carry.

To get full value out of such wave observations will require having
theoretical computations with which to compare them
\cite{hughes_flanagan}.  There is no hope to perform such computations
analytically; they can only be done as supercomputer simulations.
The development of such simulations
is a major effort within the world's relativity community.

{\bf NS/NS Coalescence:}
The final coalescence of NS/NS binaries should produce waves that are
sensitive to the equation of state of nuclear matter, so
such coalescences have the potential to teach us about the
nuclear equation of state \cite{last3minutes}.  In essence, we will be
studying nuclear physics via the collisions of atomic nuclei that have
nucleon numbers $A \sim 10^{57}$---somewhat larger than physicists
are normally
accustomed to.
The accelerator used to drive these nuclei up to the speed of light is
the binary's self gravity, and the radiation by which the details of the
collisions are probed is gravitational.

A number of research groups
\cite{nakamura_etal_nsns1,nakamura_etal_nsns2,rasio_shapiro,nakamura_nishinomiya,kochanek,bildsten_cutler,centrella,davies_melvyn}
are engaged in numerical astrophysics
simulations of NS/NS coalescence, with the goal not only to predict the
emitted gravitational waveforms and their dependence on equation of
state, but also (more immediately) to learn whether such
coalescences
might power the $\gamma$-ray bursts that have been a major astronomical
puzzle since their discovery in the early 1970s.
If advanced LIGO interferometers were now in operation, they could
report definitively whether or not the $\gamma$-bursts are produced by
NS/NS binaries; and if the answer were yes, then the combination of
$\gamma$-burst data and gravitational-wave data could bring valuable
information that neither could bring by itself.  For example, we could
determine when, to within a few msec, the $\gamma$-burst is emitted
relative to the moment the NS's first begin to touch; and by
comparing the $\gamma$ and gravitational times of arrival,
we might test whether gravitational waves propagate with
the speed of light to a fractional precision of
$\sim 0.01{\rm sec}/3\times10^9\, {\rm lyr} \sim 10^{-19}$.

Unfortunately, the final NS/NS coalescence will emit its gravitational
waves in the kHz frequency band ($800 {\rm Hz} \alt f \alt 2500 {\rm
Hz}$) where photon shot noise will prevent them from being studied by
the standard, ``workhorse,'' broad-band
interferometers of Fig.\ \ref{fig:CBStrengthSensitivity}.
However, a specially configured (``dual-recycled'')
interferometer invented by Brian Meers \cite{meers},
which could have enhanced sensitivity in the kHz
region at the price of reduced sensitivity elsewhere, may be able to
measure the waves and extract their equation of state information,
as might massive, spherical bar detectors
\cite{last3minutes,kennefick_laurence_thorne}. Such measurements will
be very difficult and are likely only when the LIGO/VIRGO
network has reached a mature stage.

{\bf NS/BH Coalescence:} A NS spiraling into a BH of mass $M \agt 10
M_\odot$ should be swallowed more or less whole.  However, if the BH is
less massive than roughly $10 M_\odot$, and especially if it is rapidly
rotating, then the NS will tidally disrupt before being swallowed.
Little is known about the disruption and accompanying waveforms.  To
model them with any reliability will likely require full numerical
relativity, since the circumferences of the BH and NS will be comparable
and their physical separation at the moment of disruption
will be of order their separation. As with NS/NS, the coalescence
waves should
carry equation of state information and will come out in the kHz band,
where their detection will require advanced, specialty detectors.

{\bf Christodoulou Memory:} As the coalescence waves depart from
their source, their energy creates (via the nonlinearity of Einstein's
field equations) a secondary wave called the ``Christodoulou memory''
\cite{christodoulou,thorne_memory,wiseman_will_memory}.  Whereas the primary
waves may have frequencies in the kHz band, the memory builds up on the
timescale of the primary energy emission profile, which is likely to be
of order 0.01 sec, corresponding to a memory frequency in the optimal
band for the LIGO/VIRGO workhorse interferometers, $\sim 100$Hz.
Unfortunately, the memory is so weak that only very advanced
interferometers have much chance of detecting and studying it---and
then, perhaps only for BH/BH coalescences and not for NS/NS or NS/BH
\cite{kennefick_memory}.

\subsection{Stellar Core Collapse and Supernovae}

Several features of the stellar core collapse, which triggers
supernovae, can produce significant gravitational radiation in the
high-frequency band. We shall
consider these features in turn, the most weakly radiating first.

\subsubsection{Boiling of the Newborn Neutron Star}

\begin{figure}
\vskip23.2pc
\special{hscale=60 vscale=60 hoffset=8 voffset=-5
psfile=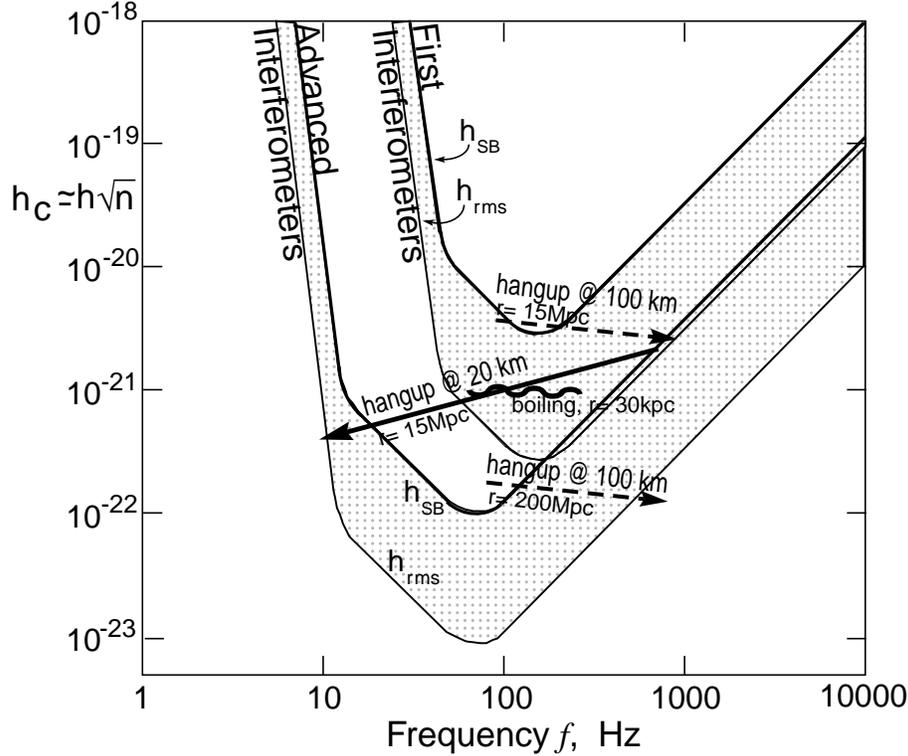}
\caption{Characteristic amplitudes of the gravitational waves from
various processes accompanying stellar core collapse and supernovae,
compared with projected sensitivities of LIGO's interferometers.
}
\label{fig:supernovae}
\end{figure}

Even if the collapse is spherical, so it cannot radiate any
gravitational waves at all, it should
produce a convectively unstable neutron
star that ``boils'' vigorously (and nonspherically) for the first
$\sim 0.1$ second of its life \cite{bethe}.  The boiling dredges
up high-temperature
nuclear matter ($T\sim 10^{12}$K) from the neutron star's central regions,
bringing it to the surface (to the ``neutrino-sphere''), where it
cools by
neutrino emission before being swept back downward and reheated.  Burrows
estimates \cite{burrows,burrows1} that the  boiling
should generate $n \sim 10$ cycles of gravitational waves with
frequency $f\sim 100$Hz and amplitude
$h \sim 3 \times 10^{-22} (30{\rm kpc}/r)$ (where $r$ is the distance to
the source), corresponding to a characteristic amplitude $h_c \simeq
h\sqrt n \sim 10^{-21} (30{\rm kpc}/r)$; cf.\ Fig.\
\ref{fig:supernovae}.  LIGO/VIRGO will be able to detect such waves only
in the local group of galaxies, where the supernova rate is probably no
larger than $\sim 1$ each 10 years.  However, neutrino detectors have a
similar range, and there could be a high scientific payoff from
correlated observations of the gravitational waves emitted by the
boiling's mass motions and neutrinos emitted from the boiling
neutrino-sphere.

\subsubsection{Axisymmetric Collapse, Bounce, and Oscillations}

Rotation will centrifugally flatten the collapsing
core, enabling it to radiate as it implodes.  If the core's angular
momentum is small enough that centrifugal forces do
not halt or strongly slow the collapse before it reaches
nuclear densities, then the core's collapse, bounce, and subsequent
oscillations are likely to be axially symmetric.  Numerical
simulations \cite{finn_collapse,monchmeyer} show that in this case the waves
from collapse, bounce, and oscillation
will be quite weak: the total energy radiated as gravitational waves
is not likely to exceed $\sim 10^{-7}$ solar masses (about 1 part in a
million of the collapse energy) and might often be much less than
this; and correspondingly, the waves'
characteristic amplitude will be $h_c \alt 3\times 10^{-21}(30{\rm
kpc}/r)$.  These collapse-and-bounce waves will come off at frequencies
$\sim 200$ Hz to $\sim 1000$ Hz, and will precede the boiling waves by a
fraction of a second.  Like the boiling waves, they probably cannot be
seen by LIGO/VIRGO beyond the local group of galaxies and thus will be a
very rare occurrence.

\subsubsection{Rotation-Induced Bars and Break-Up}

If the core's rotation is large enough to strongly flatten the
core before or as it reaches nuclear density,
then a dynamical and/or
secular instability is likely to break the core's axisymmetry.
The core will be transformed into
a bar-like configuration that spins end-over-end like
an American football, and that might even break up into two or more massive
pieces.  In this case, the radiation from the spinning bar or orbiting
pieces {\it could} be almost as strong as that from a coalescing neutron-star
binary, and thus could be seen by the LIGO/VIRGO first interferometers
out to the distance of the Virgo cluster (where the supernova rate is
several per
year) and by advanced interferometers out to several hundred Mpc
(supernova rate $\sim 10^4$ per year); cf.\ Fig.\ \ref{fig:supernovae}.
It is far
from clear what fraction of collapsing cores will have enough angular
momentum to break their axisymmetry, and what fraction of those will
actually radiate at this high rate; but even if only $\sim 1/1000$ or
$1/10^4$ do so, this could ultimately be a very interesting source for
LIGO/VIRGO.

Several specific scenarios for such non-axisymmetry have been identified:

{\bf Centrifugal hangup at $\bf \sim 100$km radius:} If the
pre-collapse core is rapidly spinning (e.g., if it is a white dwarf that
has been spun up by accretion from a companion), then the collapse may
produce a highly flattened, centrifugally supported disk with most of
its mass at radii $R\sim 100$km, which then (via instability)
may transform itself into a bar or may bifurcate.  The bar or
bifurcated lumps will radiate gravitational waves at twice their rotation
frequency, $f\sim 100$Hz --- the optimal frequency for LIGO/VIRGO
interferometers.  To shrink on down to $\sim 10$km size, this
configuration must shed most of its angular momentum.  {\it If} a
substantial fraction of the angular momentum goes into
gravitational waves, then independently of the strength of the bar,
the waves will be nearly as strong as those from a coalescing binary.
The reason is this:
The waves' amplitude $h$ is proportional to the bar's ellipticity $e$,
the number of cycles $n$ of wave emission is proportional to $1/e^2$, and the
characteristic amplitude $h_c = h\sqrt n$ is thus independent of the
ellipticity and is about the same whether the configuration is a bar or
is two lumps \cite{schutz_grg89}.  The resulting waves will thus have $h_c$
roughly half as large, at $f\sim 100$Hz, as those from a NS/NS binary
(half as large because each lump might be half as massive as a NS), and
they will chirp upward in frequency in a manner similar to those from a
binary.

It is rather likely, however, that most of
the excess angular momentum does {\it
not} go into gravitational waves, but instead goes largely into hydrodynamic
waves as the bar or lumps, acting like a propeller, stir up the
surrounding stellar mantle.  In this case, the radiation will be
correspondingly weaker.

{\bf Centrifugal hangup at $\bf R\sim 20$km:}  Lai and Shapiro
\cite{lai} have explored the case of centrifugal
hangup at radii not much lager than the final neutron star, say $R\sim
20$km.  Using compressible ellipsoidal models, they have deduced that,
after a brief period of dynamical bar-mode instability with wave
emission at $f\sim 1000$Hz (explored by
Houser, Centrella, and Smith \cite{houser}), the star switches to a secular
instability in which the bar's angular velocity gradually slows while
the material of which it is made retains its high rotation speed and
circulates through the slowing bar.  The slowing bar emits waves that sweep
{\it downward} in frequency through the LIGO/VIRGO optimal band $f\sim 100$Hz,
toward $\sim 10$Hz. The characteristic amplitude (Fig.\
\ref{fig:supernovae}) is only modestly smaller than for the upward-sweeping
waves from hangup at $R\sim 100$km, and thus such waves should be
detectable near the Virgo Cluster by the first LIGO/VIRGO interferometers,
and at distances of a few 100Mpc by advanced interferometers.

{\bf Successive fragmentations of an accreting, newborn neutron star:}
Bonnell and Pringle \cite{pringle} have focused on the evolution of the
rapidly spinning, newborn neutron star as it quickly accretes more and
more mass from the pre-supernova star's inner mantle.  If the accreting
material carries high angular momentum, it may trigger a renewed bar
formation, lump formation, wave emission, and coalescence, followed by more
accretion, bar and lump formation, wave emission, and coalescence.  Bonnell
and Pringle
speculate that hydrodynamics, not wave emission, will drive this
evolution, but that the total energy going into gravitational waves might be
as large as $\sim 10^{-3}M_\odot$.  This corresponds to $h_c \sim 10^{-21}
(10{\rm Mpc}/r)$.

\subsection{Spinning Neutron Stars; Pulsars}

As the neutron star settles down into its final state, its crust begins
to solidify (crystalize). The solid
crust will assume nearly the oblate axisymmetric shape that
centrifugal forces are trying to maintain,
with poloidal ellipticity $\epsilon_p \propto$(angular velocity of
rotation)$^2$.  However, the principal axis
of the star's moment of inertia tensor may deviate from its spin axis
by some small ``wobble angle'' $\theta_w$, and the star may
deviate slightly from axisymmetry about its principal axis; i.e., it may
have a slight ellipticity $\epsilon_e$ in its equatorial plane.

As this slightly imperfect crust spins, it will radiate gravitational
waves \cite{zimmermann}: $\epsilon_e$ radiates at twice the rotation frequency,
$f=2f_{\rm
rot}$ with
$h\propto \epsilon_e$, and the wobble angle couples to $\epsilon_p$ to
produce waves at $f=f_{\rm rot} + f_{\rm prec}$
(the precessional sideband of the rotation frequency) with amplitude
$h\propto \theta_w \epsilon_p$.  For typical neutron-star masses and
moments of inertia, the wave amplitudes are
\begin{equation}
h \sim 6\times 10^{-25} \left({f_{\rm rot}\over 500{\rm Hz}}\right)^2
\left({1{\rm kpc}\over r}\right)\left({\epsilon_e \hbox{ or }\theta_w\epsilon_p
\over 10^{-6}}\right)\;.
\label{hpulsar}
\end{equation}

The neutron star gradually spins down, due in part to gravitational-wave
emission but perhaps more strongly due to electromagnetic torques associated
with its spinning magnetic field and pulsar emission.
This spin-down reduces the strength of centrifugal forces, and thereby
causes the star's poloidal ellipticity $\epsilon_p$ to decrease, with
an accompanying breakage and resolidification of its crust's crystal structure
(a ``starquake'') \cite{starquake}.
In each starquake, $\theta_w$, $\epsilon_e$, and
$\epsilon_p$ will all change suddenly, thereby changing the amplitudes of the
star's two gravitational ``spectral lines'' $f=2f_{\rm rot}$ and
$f=f_{\rm rot} + f_{\rm prec}$.  After each quake, there should be a
healing period in which the star's fluid core and solid crust, now rotating
at different speeds, gradually regain synchronism.
By monitoring the
amplitudes, frequencies, and phases of the two gravitational-wave
spectral lines, and by
comparing with timing of
the electromagnetic pulsar emission, one might learn much about the
physics of the neutron-star interior.

How large will the quantities $\epsilon_e$ and $\theta_w \epsilon_p$ be?
Rough estimates of the crustal shear moduli and breaking strengths suggest an
upper limit in the range $\epsilon_{\rm max} \sim 10^{-4}$
to $10^{-6}$, and it might be that typical values are
far below this.  We are extremely ignorant, and
correspondingly there is much to be learned from searches for
gravitational waves from spinning neutron stars.

One can estimate the sensitivity of LIGO/VIRGO (or any other broad-band
detector)
to the periodic waves from such a source by multiplying the waves'
amplitude $h$ by the square root of the number of cycles over which one
might integrate to find the signal, $n= f \hat \tau$ where $\hat\tau$ is the
integration time.  The resulting
effective signal strength, $h\sqrt{n}$, is larger than $h$ by
\begin{equation}
\sqrt n = \sqrt{f\hat\tau} = 10^5 \left( {f\over1000{\rm Hz}}\right)^{1/2}
\left({\hat\tau\over4{\rm months}}\right)^{1/2}\;.
\label{ftau}
\end{equation}
This $h\sqrt n$  should be compared (i) to the
detector's rms broad-band noise level for sources in a random direction,
$\sqrt5 h_{\rm rms}$, to deduce a
signal-to-noise ratio, or (ii) to $h_{\rm SB}$ to deduce a sensitivity for
high-confidence detection when one does not know the waves' frequency in
advance \cite{300yrs}.
Such a comparison suggests that the first interferometers in
LIGO/VIRGO might possibly see waves from nearby spinning
neutron stars, but the odds of success are very unclear.

The deepest searches for these nearly periodic waves will be
performed by narrow-band detectors, whose sensitivities are enhanced
near some chosen frequency at the price of sensitivity loss
elsewhere---e.g., ``dual recycled'' interferometers or resonant bars.
With ``advanced-detector technology,'' dual-recycled interferometers
might be able to detect with confidence all spinning neutron stars
that have \cite{300yrs}
\begin{equation}
(\epsilon_e \hbox{ or } \theta_w\epsilon_p ) \agt 3\times10^{-10} \left(
{500 {\rm Hz}\over f_{\rm rot}}\right)^2 \left({r\over 1000{\rm pc}}\right)^2.
\label{advancedpulsar}
\end{equation}
There may well be a large number of such neutron stars in our galaxy; but
it is also conceivable that there are none.  We are extremely
ignorant.

Some cause for optimism arises from several physical mechanisms that
might generate radiating ellipticities large compared to
$3\times10^{-10}$:
\begin{itemize}

\item It may be that, inside the superconducting cores of
many neutron stars, there are trapped magnetic fields with mean
strength $B_{\rm core}\sim10^{13}$G or even
$10^{\rm 15}$G.
Because such a field is actually concentrated in flux
tubes with $B = B_{\rm crit} \sim 6\times 10^{14}$G surrounded by
field-free superconductor, its mean pressure is $p_B = B_{\rm core} B_{\rm
crit}/8\pi$. This pressure could produce a radiating ellipticity
$\epsilon_{\rm e} \sim \theta_w\epsilon_p \sim p_B/p \sim 10^{-8}B_{\rm
core}/10^{13}$G (where $p$ is the core's material pressure).

\item Accretion onto a spinning neutron star can drive precession (keeping
$\theta_w$ substantially nonzero), and thereby might produce measurably
strong
waves \cite{schutz95}.

\item If a neutron star is born rotating very rapidly,
then it may experience a
gravitational-radiation-reaction-driven instability.  In this
``CFS'' (Chandrasekhar, \cite{cfs_chandra} Friedman, Schutz
\cite{cfs_friedman_schutz}) instability,
density waves propagate around the
star in the opposite direction to its rotation, but are dragged forward
by the rotation.  These density waves produce gravitational waves that
carry positive energy as seen by observers far from the star, but
negative energy from the star's viewpoint; and because the
star thinks it is losing negative energy, its density waves get
amplified.  This intriguing mechanism is similar to that by which
spiral density waves are produced in galaxies.  Although the CFS
instability was once thought ubiquitous for spinning stars, we now
know that neutron star viscosity will kill it, stabilizing the star and
turning off the waves, when the star's temperature is above some limit
$\sim 10^{10}{\rm K}$ \cite{cfs_lindblom}
and below some limit $\sim 10^9 {\rm K}$ \cite{cfs_mendell_lindblom}; and
correspondingly,
the instability
should operate only during the first few years of a neutron
star's life, when $10^9 {\rm K} \alt T \alt 10^{10}\rm K$.

\end{itemize}

\section{LISA: The Laser Interferometer Space Antenna}

Turn, now, from the high-frequency band, 1---$10^4$ Hz,
to the low-frequency band, $10^{-4}$---1 Hz.  At present, the most
sensitive
gravitational-wave searches at low frequencies are those carried out by
researchers at NASA's Jet propulsion Laboratory, using microwave-frequency
Doppler tracking of interplanetary spacecraft. These
searches are done at rather low cost, piggy-back on missions
designed for other purposes.  Although they have a definite possibility
of success, the odds are against them.  Their best past
sensitivities to bursts, for
example, have been $h_{\rm SB} \sim 10^{-14}$, and prospects are good
for reaching $\sim 10^{-15}$---$10^{-16}$ in the next 5 to 10 years.  However,
the strongest low-frequency bursts
arriving several times per year might be no larger than
$\sim 10^{-18}$; and the domain of an assured plethora of signals is
$h_{\rm SB} \sim 10^{-19}$---$10^{-20}$.

In the 2014 time frame,
the European Space Agency (ESA) and/or NASA is likely to fly a {\it
Laser Interferometer Space Antenna} (LISA) which will achieve $h_{\rm
SB} \alt 10^{-20}$ over the frequency band $3\times 10^{-4}{\rm Hz} \alt
f \alt 3\times10^{-2} {\rm Hz}$.

\subsection{Mission Status}

LISA is largely an outgrowth of 15 years of studies by Peter
Bender and colleagues at the University of Colorado.
Unfortunately, the prospects for NASA to fly such a mission have not
looked good in the early 1990s.  By contrast, prospects in Europe have looked
much better, so a largely European consortium was put
together in 1993 with Bender's participation but under the leadership of
Karsten Danzmann (Hannover) and
James Hough (Glasgow), to propose LISA to the European Space Agency.
This
proposal has met with considerable success; LISA might well achieve
approval to fly as an ESA Cornerstone Mission around 2014
\cite{cornerstone}.
Members of the American gravitation community hope that NASA will
join together with ESA in this endeavor, and that working jointly, ESA
and NASA will be able to fly LISA considerably sooner than 2014.

\begin{figure}
\vskip13pc
\special{hscale=60 vscale=60 hoffset=40 voffset=2
psfile=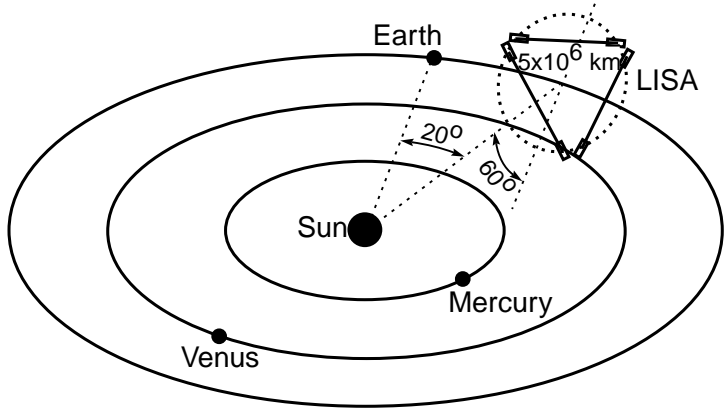}
\caption{LISA's orbital configuration.
}
\label{fig:lisa_orbit}
\end{figure}

\subsection{Mission Configuration}

As presently conceived, LISA will consist of six compact, drag-free
spacecraft (i.e. spacecraft that are shielded from buffeting by solar
wind and radiation pressure, and that thus move very nearly on geodesics of
spacetime).  All six spacecraft would be launched simultaneously by a
single Ariane rocket. They
would be placed into the same heliocentric orbit as the Earth
occupies, but would follow 20$^{\rm o}$ behind the Earth; cf.\ Fig.\
\ref{fig:lisa_orbit}.  The spacecraft would fly in pairs, with each pair
at the vertex of an equilateral triangle that is inclined at an angle of
60$^{\rm o}$ to the Earth's orbital plane. The triangle's arm length would be 5
million km ($10^6$ times larger than LIGO's arms!).  The six spacecraft would
track each other optically, using one-Watt YAG laser beams.  Because of
diffraction
losses over the $5\times10^6$km arm length, it is not feasible to
reflect the beams back and forth between mirrors as is done with LIGO.
Instead, each spacecraft will have its own laser; and the lasers will be
phase locked to each other, thereby achieving the same kind of
phase-coherent out-and-back light travel as LIGO achieves with mirrors.
The six-laser, six-spacecraft configuration thereby functions as three,
partially independent but partially redundant,
gravitational-wave interferometers.

\subsection{Noise and Sensitivity}

\begin{figure}
\vskip23.5pc
\special{hscale=60 vscale=60 hoffset=14 voffset=-412
psfile=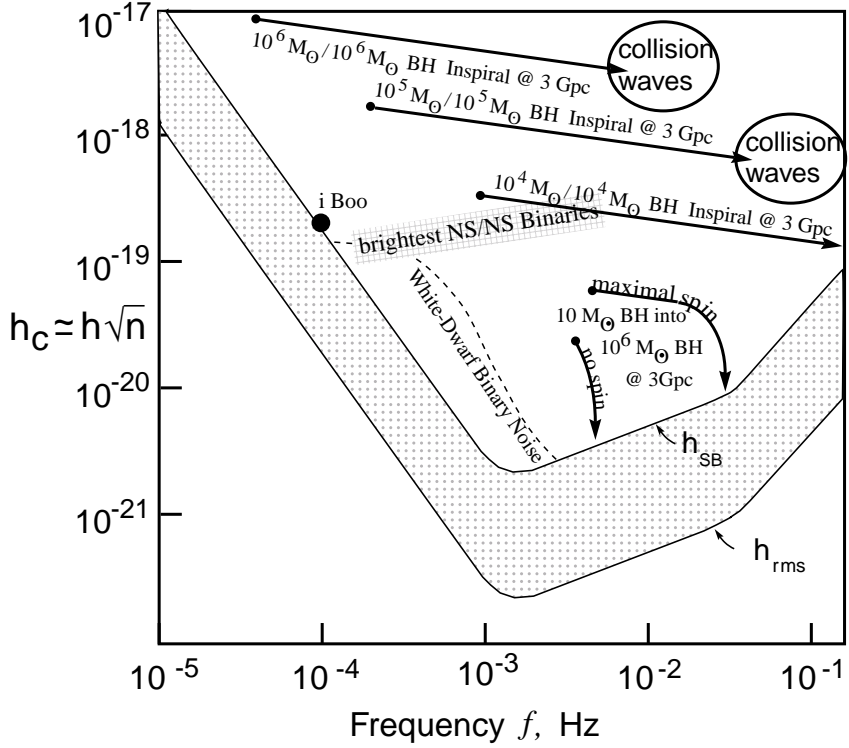}
\caption{LISA's projected broad-band noise $h_{\rm rms}$ and sensitivity
to bursts $h_{\rm SB}$, compared with the strengths of the waves from
several low-frequency sources. [{\it Note:} When members of the LISA team
plot curves analogous to this, they show the sensitivity curve (top of
stippled region) in units of the amplitude of a periodic signal
that can be detected with $S/N=5$ in one year of integration; that
sensitivity to periodic sources is related
to the $h_{\rm SB}$ used here by $h_{\rm SP} = h_{\rm SB}/\protect\sqrt{f\cdot
3\times 10^7\rm{sec}}$.]
}
\label{fig:lisa_noise}
\end{figure}

Figure \ref{fig:lisa_noise} depicts the expected noise and sensitivity of
LISA in the same language as we have used for LIGO (Fig.\
\ref{fig:CBStrengthSensitivity}).
The curve at the bottom of the stippled region
is $h_{\rm rms}$, the rms noise, in a bandwidth equal to frequency,
for waves with optimum direction and polarization.  The top of the
stippled region is $h_{\rm SB} = 5\sqrt5 h_{\rm rms}$, the sensitivity
for high-confidence detection ($S/N=5$) of a broad-band burst coming from
a random direction, assuming Gaussian noise.

At frequencies $f\agt 10^{-3}$Hz, LISA's noise is due to photon counting
statistics (shot noise).  The noise curve steepens at $f\sim
3\times10^{-2}$Hz because at larger $f$ than that, the waves' period is
shorter than the round-trip light travel time in one of LISA's arms.
Below $10^{-3}$Hz, the noise is due to buffeting-induced random motions
of the spacecraft
that are not being properly removed by the drag-compensation system.
Notice that, in terms of dimensionless amplitude, LISA's sensitivity is
roughly the same as that of LIGO's first interferometers (Fig.\
\ref{fig:CBStrengthSensitivity}),
but at 100,000 times lower frequency.  Since
the waves' energy flux scales as $f^2 h^2$, this corresponds to $10^{10}$
better energy sensitivity than LIGO.

\subsection{Observational Strategy}

LISA can detect and study, simultaneously, a wide variety of different
sources scattered over all directions on the sky.  The key to
distinguishing the different sources is the different time evolution of
their waveforms.  The key to determining each source's direction, and
confirming that it is real and not just noise, is the manner in which
its waves' amplitude and frequency are modulated by LISA's complicated
orbital motion---a motion in which the interferometer
triangle rotates around
its center once per year, and the interferometer plane rotates
around the normal to the Earth's orbit once per year.  Most sources will
be observed for a year or longer, thereby making full use of these
modulations.

\section{Low-Frequency Gravitational-Wave Sources}

\subsection{Waves from Binary Stars}

LISA has a large class of guaranteed sources: short-period binary stars
in our
own galaxy.  A specific example is the classic binary 44 i Boo
(HD133640),
a $1.35 M_\odot$/
$0.68 M_\odot$ system just 12 parsecs from Earth, whose wave
frequency $f$ and characteristic amplitude $h_c = h\sqrt n$
are depicted in Fig.\ \ref{fig:lisa_noise}.  (Here
$h$ is the waves' actual amplitude and $n = f\hat\tau$ is the number of
wave cycles during $\hat\tau =$1 year of signal integration.)
Since 44 i Boo lies right on the $h_{\rm SB}$ curve, its signal to
noise ratio in one year of integration should be $S/N = 5$.

To have an especially short period, a binary must be made of especially
compact bodies---white dwarfs (WD), neutron stars (NS), and/or black
holes (BH).  WD/WD binaries are thought to be so numerous that they
might produce a stochastic background of gravitational waves, at the level
shown in Fig.\ \ref{fig:lisa_noise}, that will hide some other interesting
waves
from view \cite{wdbinaries}.  Since WD/WD binaries are very dim optically,
their actual numbers are not known for sure; Fig.\ \ref{fig:lisa_noise}
might be an overestimate.

Assuming a NS/NS coalescence rate of 1 each $10^5$ years in our galaxy
\cite{phinney,narayan}, the shortest period NS/NS binary should have a
remaining life of about $5\times 10^4$ years, corresponding to a
gravitational-wave frequency today of $f\simeq 5\times10^{-3}$Hz, an
amplitude (at about $10$kpc distance) $h \simeq 4 \times 10^{-22}$, and a
characteristic amplitude (with one year of integration time) $h_c \simeq
2\times 10^{-19}$.  This is depicted in Fig.\ \ref{fig:lisa_noise} at the
right edge of the region marked ``brightest NS/NS binaries''.  These
brightest NS/NS binaries can be studied by LISA with
the impressive signal to noise ratios $S/N \sim 50$ to $500$.

\subsection{Waves from the Coalescence of Massive Black Holes in
Distant Galaxies}

LISA would be a powerful instrument for studying massive black holes in
distant galaxies.  Figure \ref{fig:lisa_noise} shows, as examples, the
waves from several massive black hole binaries at 3Gpc distance.
The waves sweep upward in frequency
(rightward in the diagram) as the holes spiral together.  The black dots
show the waves' frequency one year before the holes' final collision and
coalescence, and the arrowed lines show the sweep of frequency and
characteristic amplitude $h_c = h\sqrt n$ during that last year.  For
simplicity, the figure is restricted to binaries with equal-mass black
holes:
$10^4M_\odot / 10^4 M_\odot$, $10^5M_\odot / 10^5 M_\odot$,
and $10^6M_\odot / 10^6 M_\odot$.

By extrapolation from these three examples, we see that LISA can
study much of the last year of inspiral, and the waves
from the final collision and coalescence,
whenever the holes' masses are in the range $3\times 10^4 M_\odot
\alt M \alt 10^8
M_\odot$.  Moreover, LISA can study the final coalescences with remarkable
signal to noise ratios: $S/N \agt 1000$.
Since these are much larger $S/N$'s than LIGO/VIRGO is likely to achieve,
we can expect LISA to refine the experimental understanding of black-hole
physics, and of highly nonlinear vibrations of warped spacetime,
which LIGO/VIRGO initiates---{\it provided} the rate of massive
black-hole coalescences is of order one per
year in the Universe or higher.  The rate might well be that high, but
it also might be much lower.

By extrapolating Fig.\ \ref{fig:lisa_noise} to lower BH/BH masses, we
see that LISA can observe the last few years of inspiral, but not the
final collisions, of binary black holes in the range
$100M_\odot \alt M \alt 10^4 M_\odot$, out to cosmological distances.

Extrapolating the BH/BH curves to lower frequencies using the
formula (time to final coalescence$)\propto f^{-8/3}$, we see that
equal-mass BH/BH binaries enter LISA's frequency band roughly 1000 years
before their final coalescences, more or less independently of their
masses, for the range $100 M_\odot \alt M \alt 10^6 M_\odot$.  Thus, if the
coalescence rate were to turn out to be one per year, LISA would see
roughly 1000 additional massive binaries that are slowly spiraling
inward, with inspiral rates $df/dt$ readily measurable.  From the inspiral
rates, the amplitudes of the two polarizations, and the waves' harmonic
content, LISA can determine each such binary's luminosity distance,
redshifted chirp mass $(1+z) M_c$, orbital inclination,
and eccentricity; and from the waves' modulation by LISA's orbital
motion, LISA can learn the direction to the binary with an accuracy of
some tens of arcminutes.

\subsection{Waves from Compact Bodies Spiraling into Massive Black
Holes in Distant Galaxies}
\label{inspiral_waves}

When a compact body with mass $\mu$ spirals into a much more massive black
hole with mass $M$, the body's orbital energy $E$ at fixed frequency
$f$ (and correspondingly at fixed orbital radius $a$)
scales as $E \propto \mu$,
the gravitational-wave luminosity $\dot E$ scales as
$\dot E \propto \mu^2$, and the time to
final coalescence thus scales as $t \sim E/\dot E \propto 1/\mu$.  This
means that the smaller is $\mu/M$,
the more orbits are spent in the hole's strong-gravity region, $a\alt
10GM/c^2$, and thus the more detailed and accurate will be the map of the
hole's spacetime geometry, which is encoded in the emitted waves.

For holes observed by LIGO/VIRGO, the most extreme mass ratio that we
can hope for is $\mu/M \sim 1M_\odot/300 M_\odot$, since for $M>300M_\odot$ the
inspiral waves are pushed to frequencies below the LIGO/VIRGO band.
This limit on $\mu/M$ seriously constrains the accuracy with which
LIGO/VIRGO can hope to map out the spacetime geometries of black
holes and test the black-hole no-hair theorem (Sec. \ref{coalescing_binaries}).
By contrast, LISA can observe the final inspiral waves from objects of
any mass $M\agt 0.5M_\odot$ spiraling into holes of mass $3\times 10^5 M_\odot
\alt M \alt 3\times10^7M_\odot$.

Figure \ref{fig:lisa_noise} shows the
example of a $10M_\odot$ black hole spiraling into a $10^6M_\odot$ hole
at 3Gpc distance.  The inspiral orbit and waves are strongly influenced
by the hole's spin.  Two cases are shown \cite{finn_thorne}:
an inspiraling circular orbit
around a non-spinning hole, and a prograde, circular, equatorial orbit
around a maximally spinning hole.
In each case the dot at the upper left end of the
arrowed curve is the frequency and characteristic amplitude one year
before the final coalescence.  In the nonspinning case, the small hole
spends its last year spiraling inward from $r\simeq 7.4 GM/c^2$
(3.7 Schwarzschild
radii) to its last stable circular orbit at $r=6GM/c^2$ (3 Schwarzschild
radii).  In the maximal spin case, the last year is spent traveling from
$r=6GM/c^2$ (3 Schwarzschild radii) to the last stable orbit at $r=GM/c^2$
(half a
Schwarzschild radius).  The $\sim 10^5$ cycles of waves during this last
year should carry, encoded in themselves, rather accurate values for
the massive hole's lowest few multipole moments \cite{ryan}.  If the
measured moments satisfy the ``no-hair'' theorem (i.e., if they are all
determined uniquely by the measured mass and spin in the manner of the
Kerr metric), then we can be sure the central body is a black hole.  If
they violate the no-hair theorem, then (assuming general relativity is
correct), either the central body was not a black hole, or an accretion
disk or other material was perturbing its orbit \cite{chakrabarti}.
{}From the evolution of the waves one can hope to determine which is
the case, and to explore the properties of the central body and its
environment.

Models of galactic nuclei, where massive holes reside, suggest that
inspiraling stars and small holes typically will be in rather eccentric
orbits \cite{hils_bender}.  This is because they get injected into such
orbits via
gravitational deflections off other stars, and by the time gravitational
radiation reaction becomes the dominant orbital driving force, there is
not enough inspiral left to fully circularize their orbits.  Such orbital
eccentricity will complicate the waveforms and complicate the extraction
of information from them.  Efforts to understand the emitted waveforms
are just now getting underway.

The event rates for inspiral into massive black holes are not at all
well understood.  However, since a significant fraction of all galactic
nuclei are thought to contain massive holes, and since white dwarfs and
neutron stars, as well as small black holes, can withstand tidal
disruption as they plunge toward the massive hole's horizon, and since
LISA can see inspiraling bodies as small as $\sim 0.5 M_\odot$ out to
3Gpc distance, the event rate is likely to be interestingly large.

\section{Conclusion}

It is now 35 years since Joseph Weber initiated his pioneering
development of gravitational-wave detectors \cite{weber} and 25 years
since Forward and Weiss initiated work on interferometric detectors.
Since then, hundreds of
talented experimental physicists have struggled to improve the
sensitivities of these instruments.  At last, success is in sight.  If
the source estimates described in this lecture are approximately
correct, then the planned interferometers should detect the first waves
in 2001 or several years thereafter, thereby opening up this rich new
window onto the universe.  One payoff should be deep new insights
into compact astrophysical bodies and their roles in binary systems.

\section{Acknowledgments}

For insights into the rates of coalescence of compact binaries, I thank
Sterl Phinney.  My group's research on gravitational waves from compact
bodies and the waves' relevance to LIGO/VIRGO and LISA is supported in part
by NSF grants AST-9417371 and PHY-9424337 and by NASA grant NAGW-4268.
This written version of my lecture has been adapted from a longer review
article that I am
writing for the Proceedings of the Snowmass '94 Summer Study on
Particle and Nuclear Astrophysics and Cosmology in the Next Millenium.


\end{document}